# On the Impact of Energy Dissipation Model on Characteristic Distance in Wireless Networks

Ashraf Hossain

*Department of Electronics & Communication Engg., Aliah University, DN-47, Sector-V, Salt Lake City, Kolkata-91, India*
E.mail: ashraf.ece@aliah.ac.in; hossain_ashraf@rediffmail.com

*Abstract–* **In this paper we investigate the dependency of characteristic distance on energy dissipation model. Both the many-to-one and any-to-any communication paradigm have been presented for performance analysis. Characteristic distance has been derived for three different cases. This study will be useful for designing an energy-efficient wireless network where nodes are energy-constrained.**

## I. INTRODUCTION

In wireless networks e.g. mobile ad-hoc network, wireless sensor network where nodes are scattered over an area of interest for specific purposes. Some times nodes are deployed regularly in a linear fashion. The path loss of radio communication varies with distance in a greater-than-linear fashion [1]. Source node can send data to the sink using single hop or multi-hop. There is a trade-off between energy consumption and mode of communication. The best mode of communication depends on the path loss exponent, radio parameters, link distance etc. If the link distance is very small then no need to introduce multi-hop mode of communication. On the other hand, for a longer link multi-hop mode of communication is suitable. The hops that are very small lead to excessive receive energy. The hops that are too large lead to excessive transmit energy. The characteristic distance is the optimum one in between these two extremes [1].

Bhardwaj *et al.* [2] have found the characteristic distance for any-to-any linear network where one source node sends information to the sink using optimal number of relay nodes. Gao *et al.* [1] have found the characteristic distance for any-to-any linear network while additionally considering the idling state energy dissipation. Shelby *et al.* [3] have also found the optimal spacing for a many-to-one linear link.

In this paper, we present the impact of energy dissipation model on characteristic distance. We present characteristic distance for three different cases. The remaining of the paper is organized as follows. Section II describes the system model. Section III presents the characteristic distance with some numerical examples. Finally, section IV concludes the paper.

## II. SYSTEM MODEL

A linear array of $K$ wireless nodes is considered with the sink at one end (Fig. 1). We assume that all the $K$ nodes have same initial energy of $E_0$ units. Also we assume that the sink node is not energy-constrained. The distance between $i^{th}$ node and $(i–1)^{th}$ node is indicated as $h_i$ units for $2 \le i \le K$. The distance between the 1st node and the sink is denoted as $h_1$. The farthest $K^{th}$ node is at a distance of $D$ units from the sink.

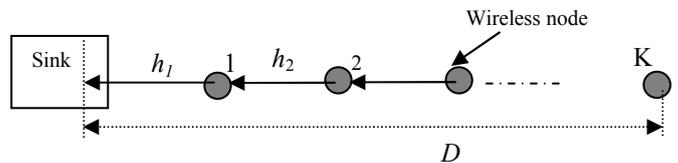

Figure 1. Linear wireless network.

For this model,

$$\sum_{i=1}^{K} h_i = D \qquad (1)$$

We assume a linear network in the context of mobile ad-hoc network (MANET) and wireless sensor network (WSN). The communication paradigm for MANET is any-to-any while that for WSN is many-to-one.

*A. Any-to-any communication link*

We consider a linear array of $K$ nodes over a distance $D$ (Fig. 1) where only the $K^{th}$ node sends $A$ number of packets to the sink using multi-hop mode of communication. The nearest neighbour routing towards the sink has been assumed.

*B. Many-to-one communication link*

We consider a data-gathering network where each node generates one packet of equal length ($B$ bits) over a data gathering cycle of $T_d$ second. A node sends a packet to the sink by using the nearest neighbour towards the sink as a repeater. Nodes closer to the sink are expected to forward all the packets towards the sink. No data aggregation is assumed at any node. We assume that each node can deal with $P$ packets/second. This implies that $P.T_d \ge K$.

*C. Energy dissipation model of a node*

The energy dissipation model for radio communication is assumed similar to [4], following which the energy consumed by a node for transmitting a packet over a distance $h_i$ is $E_t = e_t + e_d h_i^n$. Here $e_t$ is the amount of energy spent per packet in the transmitter electronics circuitry and $e_d h_i^n$ is the amount of energy necessary for transmitting a packet satisfactorily to the $(i–1)^{th}$ node. The constant '$e_d$' is dependent on the transmit amplifier efficiency, antenna gains and other system parameters. The path loss exponent is $n$ (usually $2.0 \leq n \leq 4.0$) [5]. On the receiving end, the amount of energy spent to capture an incoming packet of $B$ bits is $e_r$ units. The radio is assumed to consume energy even during idle state, i.e., when the radio neither transmits nor receives [1]. The idle state energy is equal to $e_{id}T_{id}P$, where $T_{id}$ is the idle time and $e_{id} = c.e_r$ is the idle state energy spent per packet duration, where $0 < c \leq 1.0$ [1]. $P$ is the packet dealing rate of the node. Perfect power control is assumed.

### III. ANALYSIS AND NUMERICAL EXAMPLES

In this section we present the characteristic distance for three different cases.

*Case-I: Any-to-any communication paradigm and node energy dissipation model similar to [4]*

Energy dissipated by the network can be expressed as

$$E = \sum_{i=1}^{K} \{(e_t + e_d h_i^n) + e_r\} A - e_r A \tag{2}$$

Equation (2) is a convex function of $h_i$. Applying Jension inequality, (2) becomes minimum when all the inter-node distances are made equal to $D/K$. Substituting $h_i$ by $D/K$ we get from (2)

$$E = KA \left[ e_t + e_r + e_d \left( \frac{D}{K} \right)^n \right] - e_r A \tag{3}$$

Now, setting $dE/dK=0$ gives the characteristic distance

$$d_{char1} = \sqrt[n]{\frac{e_t + e_r}{e_d (n-1)}} \tag{4}$$

The characteristic distance, $d_{char1}$ is independent on the number of packet, $A$. It depends on the radio parameters ($e_t$, $e_r$, $e_d$) and the path loss exponent ($n$). Equation (4) is similar to the result obtained by Bhardwaj *et al.* [2].

*Case-II: Any-to-any communication paradigm and node energy dissipation model similar to [1]*

We consider here that $K^{th}$ node sends $A$ packets over the time duration $T_d$. We assume that node dissipates energy for idling state also. Energy dissipated by the network can be expressed as

$$E = \sum_{i=1}^{K} \{(e_t + e_d h_i^n) + e_r\} A - e_r A + \sum_{i=1}^{K} e_{id}(PT_d - 2A) + e_{id} A \tag{5}$$

Applying the same argument as applied to Case-I, (5) can be simplified as

$$E = KA \left[ e_t + e_r + e_d \left( \frac{D}{K} \right)^n \right] - e_r A + K e_{id}(PT_d - 2A) + e_{id} A \tag{6}$$

Again, $dE/dK=0$ gives the characteristic distance

$$d_{char2} = \sqrt[n]{\frac{(e_t + e_r)A + e_{id}(PT_d - 2A)}{e_d (n-1) A}} \tag{7}$$

The characteristic distance, $d_{char2}$ is dependent on the number of packets, $A$. The more is the number of packets the less is the idling energy. It also depends on the radio parameters ($e_t$, $e_r$, $e_d$) and the path loss exponent ($n$). Equation (7) is similar to the result obtained by Gao *et al.* [1]. $d_{char2}$ is equal to $d_{char1}$ when $PT_d = 2A$ i.e. only when the radio is always busy for transmitting or receiving packets.

*Case-III: Many-to-one communication paradigm*

According to the system model, the number of packets received by the $i^{th}$ node per data gathering cycle is

$$A_r(i) = K - i, \text{ for } 1 \leq i \leq K \tag{8}$$

The number of packets transmitted by the $i^{th}$ node including its own packet per data gathering cycle is

$$A_t(i) = (K - i) + 1 = A_r(i) + 1, \text{ for } 1 \leq i \leq K \tag{9}$$

The duration of time the sensor node is idle over a single data gathering cycle may be expressed as:

$$T_{id}(i) = \left( T_d - \frac{2(K-i)+1}{P} \right), \text{ for } 1 \leq i \leq K \tag{10}$$

Following the energy consumption model (section-II), the total amount of energy spent by the $i^{th}$ node per data gathering cycle is,

$$E(i) = E_1 - iE_2 + e_d (K - i + 1) h_i^n, \text{ for } 1 \leq i \leq K \tag{11}$$

where, $\quad E_1 = e_t(K+1) + e_r K + e_{id}(PT_d - 2K - 1) \tag{12}$

and $\quad E_2 = (e_t + e_r - 2e_{id}) \tag{13}$

The total energy $E_{tot}$ consumed by the whole network over a single data gathering cycle can be expressed as:

$$E_{tot} = \sum_{i=1}^{K} E(i) \tag{14}$$

We are interested in finding the optimal spacing for the multi-hop many-to-one communication paradigm case. We also denote the optimal spacing as the characteristic distance for many-to-one case. Now the problem is to minimize $E_{tot}$ with the constraint $D = \sum_{i=1}^{K} h_i$. To solve this problem using method of Lagrange's multipliers, the Lagrangian $L(h_i, \lambda)$ is

$$L(h_i, \lambda) = E_{tot} + \lambda \left( D - \sum_{i=1}^{K} h_i \right) \tag{15}$$

where, $\lambda$ is the Lagrange's multiplier.
Taking partial derivatives of $L(h_i, \lambda)$ with respect to $h_i$ and equating to 0 gives

$$\frac{\partial L}{\partial h_i} = (K - i + 1) n e_d h_i^{n-1} - \lambda = 0 \tag{16}$$

Equation (16) reduces to

$$h_i = \sqrt[(n-1)]{\left( \frac{\lambda}{n e_d (K - i + 1)} \right)} \tag{17}$$

Using the constraint $\sum_{i=1}^{K} h_i = D$ we get from (17)

$$h_i = \frac{D}{\sum_{i=1}^{K} \left( \frac{1}{i} \right)^{\frac{1}{n-1}}} \left( \frac{1}{K - i + 1} \right)^{\frac{1}{n-1}} \tag{18}$$

Equation (18) is the expression for characteristic distance. It depends on the link distance ($D$), number of nodes ($K$), position of the node and the path loss exponent ($n$). It does not depend on the radio parameters. Equation (18) is similar to the result obtained by Shelby *et al.* in [3]. However, different radio energy dissipation model has been considered for the derivation.

The system parameters for numerical evaluation is presented in Table-I. Fig. 2 shows the variation of the characteristic distance with number of packets. This result corresponds to (7). It is also clear from Fig. 2 that when idle time is zero (when $PT_d = 2A$) then the characteristic distance, $d_{char2}$ is equal to 31.62 m which corresponds to the value of $d_{char1}$ in (4). It also tells that when the number of packets is more then smaller is the characteristic distance.

TABLE I
SYSTEM PARAMETERS FOR PERFORMANCE ANALYSIS

| Parameter | Value |
|---|---|
| $e_t$ | 25.6 μJ/packet |
| $e_r$ | 25.6 μJ/packet |
| $e_d$ (for n = 2) | 51.2 nJ/packet/m$^2$ |
| $e_{id}$ ($c.e_r$) | 23.04 μJ/packet |
| $T_d$ | 60 s |
| Packet dealing rate, $P$ | 1 packet/s |
| Path loss exponent, $n$ | 2.0 |

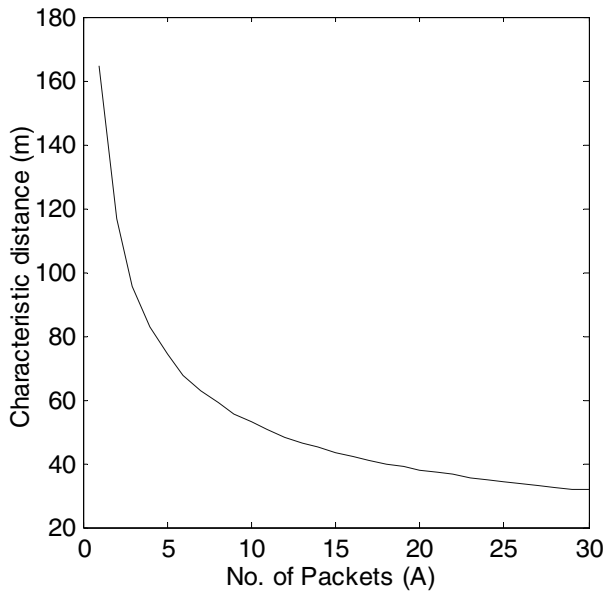

Figure 2. Variation of characteristic distance with number of packets.

## IV. CONCLUSION

The characteristic distance for the wireless networks (mobile ad-hoc network and wireless sensor network) have been studied. We have investigated the dependency of characteristic distance on energy dissipation model. In a wireless network, if a routing path is formed by the relaying nodes with a separation equal to characteristic distance then the path will dissipate least energy. This study has an importance in designing energy-efficient wireless network.